\documentclass{article_saj}
\usepackage{graphicx,saj,multicol,subeqnarray}
\usepackage{subfig}
\usepackage{natbib}
\usepackage{float}
\usepackage{xcolor}
\usepackage{widetext}
\usepackage{url}
\usepackage{amsmath,upgreek}
\usepackage[normalem]{ulem}
\definecolor{xlinkcolor}{cmyk}{1,0.6,0,0}
\usepackage[bookmarks=false,         
     pdfnewwindow=true,      
     colorlinks=true,    
     linkcolor=xlinkcolor,     
     citecolor=xlinkcolor,     
     filecolor=xlinkcolor,  
     urlcolor=xlinkcolor,      
final=true
]{hyperref}

\def\papertype{\ \hfill\ Editorial}

\def\udc{52}
\begin{document}
\parindent=.5cm
\baselineskip=3.1truemm
\columnsep=.5truecm
\newenvironment{lefteqnarray}{\arraycolsep=0pt\begin{eqnarray}}
{\end{eqnarray}\protect\aftergroup\ignorespaces}
\newenvironment{lefteqnarray*}{\arraycolsep=0pt\begin{eqnarray*}}
{\end{eqnarray*}\protect\aftergroup\ignorespaces}
\newenvironment{leftsubeqnarray}{\arraycolsep=0pt\begin{subeqnarray}}
{\end{subeqnarray}\protect\aftergroup\ignorespaces}
%


\markboth{\eightrm {\tt BVRI} PHOTOMETRIC CALIBRATION OF THE  NEDELJKOVI\'C TELESCOPE} 
{\eightrm A. VUDRAGOVI\'C {\lowercase{\eightit{et al.}}}}

\begin{strip}

{\ }

\vskip-1cm

\publ

\type

{\ }


\title{$BVRI$ PHOTOMETRIC CALIBRATION OF THE  NEDELJKOVI\'C TELESCOPE}


\authors{A. Vudragovi\'c and M. I. Jurkovic}

\vskip3mm


\address{Astronomical Observatory, Volgina 7, 11060 Belgrade, Serbia}


\Email{ana@aob.rs}


\dates{April 1, 2021}{June 1, 2021}


\summary{We have done photometric calibration of the 60 cm Nedeljkovi\'{c} telescope equipped with FLI PL 230 CCD camera, mounted at the Astronomical Station Vidojevica (Serbia), using standard stars from the Landolt's catalog. We have imaged 31 fields of standard stars using Johnson's $BVRI$ filters during three nights in August 2019. We have measured both extinction and color correction. Relating our calibrated magnitudes to the magnitudes of the standard stars from the Landolt's catalog, we have achieved accuracy of 2\%-5\% for the $BVRI$ magnitudes.}



\end{strip}

\tenrm


\section{INTRODUCTION}
\label{introduction}
\indent

The Astronomical Station Vidojevica\footnote{\url{www.vidojevica.aob.rs}} (ASV) hosts several telescopes used for various observing projects led by the Astronomical Observatory in Belgrade. It is located on the Vidojevica mountain top in the south of Serbia. For more information on the ongoing observing projects see \citet{Vince_2014}. The ASV has the following telescopes:

\indent
\parindent=.7cm
\par\hang\textindent{o} Milankovi\'c (1.4~m) telescope,
\par\hang\textindent{o} Nedeljkovi\'c (60~cm) telescope, and 
\par\hang\textindent{o} a MEADE 40~cm telescope.
\vspace{0.2cm}

The 60~cm Nedeljkovi\'c telescope (Cassegrain type) has been in use since 2011 \citep{Vince_2012}. In this article we describe the photometric calibration carried out on the this telescope during 2019 for the purpose of introducing a new observational project: the observation of pulsating stars. The CCD camera on the telescope is FLI PL 230.

In order to calibrate magnitudes and colors of any object, one needs to observe many standard stars fields, well distributed on the sky at different time during the night, making sure their colors cover the range of the target's color. These standard stars are used to define our photometric system through a set of calibration equations that relate instrumental and calibrated magnitudes. Solving a set of calibration equations determines calibration coefficients, that transform instrumental magnitudes into the calibrated magnitudes. A detailed description of the photometric procedures applicable in astronomical measurements is given in \citep{Henden}.

In this paper we describe the selection of the standard stars used to preform the photometric calibration in the Section \ref{observations} In the Section \ref{datareduction}, we provide description of the data reduction procedure used. Photometric calibration is introduced in the Section \ref{calibration}, where two different models (approaches) are discussed. Finally, we test these different models and provide the relevant coefficients for the photometric transformation to the standard Landolt's photometric system for the $BVRI$ filters in the Section \ref{results}.

\section{OBSERVATIONS}
\label{observations}
\indent

The observations were done on the $5^{th}$, $6^{th}$ and $7^{th}$ of August, 2019. The individually observed photometric standards were taken from the \citep{Landolt}. The Landolt's catalog contains 258 standard stars in 156 fields each having a $6.8' \times 6.8'$ field of view. Field of view of our instrumentation was $17.7' \times 17.8'$. Selected Landolt's fields with their celestial coordinates are listed in the Table~\ref{tab:Landolt_obs}, grouped by the date at which they were observed. 
\begin{table}[!]
\caption{Observational log of the Landolt's fields: field name is given in the first column, followed by its right ascension and declination. Fields are listed in the order of the time of observations, divided into three nights.}
\centerline{\begin{tabular}{ccc}
\hline\hline
Field & RA[h:m:s] & DEC[d:m:s] \\ \hline\hline
& $5^{th}$ August, 2019  &  \\ \hline\hline
GD2 & 00:00:34 & 33:18:41 \\
PG1648+536 & 16:16:01 & 53:29:36 \\
GD 363 & 17:17:38 & 41:53:37 \\
GD 378 & 18:18:41 & 41:05:36 \\
GD 391 & 20:20:51 & 39:15:56 \\
KUV 433-03 & 16:16:27 & 35:00:18 \\
PG1430+427 & 14:14:33 & 42:31:43 \\
SA 35 SF1 & 15:15:51 & 44:32:01 \\
SA 38 SF1 & 18:18:36 & 45:10:55 \\
SA 38 SF4 & 18:18:39 & 45:08:29 \\
SA 41 SF1 & 21:21:41 & 45:35:08 \\
SA 41 SF2 & 21:21:20 & 45:14:48 \\
GD 336 & 14:14:54 & 37:06:26 \\ \hline
\hline
& $6^{th}$ August, 2019 & \\
\hline\hline
GD 13 & 01:01:41 & 42:27:52 \\
GD 275 & 01:01:59 & 52:27:40 \\
GD 277 & 01:01:27 & 51:08:24 \\
GD 278 & 01:01:02 & 53:20:59 \\
GD 279 & 01:01:02 & 47:00:48 \\
GD 405 & 23:23:45 & 47:26:57 \\
GD 421 & 01:01:09 & 67:42:27 \\
SA 23 SF1 & 03:03:24 & 45:06:40 \\
SA 23 SF2 & 03:03:42 & 45:20:28 \\
\hline\hline
& $7^{th}$ August, 2019 & \\
\hline\hline
GD 10 & 01:01:00 & 39:31:01 \\
GD 325 & 13:13:15 & 48:29:56 \\
GD 8 & 00:00:45 & 31:34:36 \\
PG1343+578 & 13:13:02 & 57:31:37 \\
SA 20 SF1 & 00:00:48 & 45:49:51 \\
SA 20 SF2 & 00:00:25 & 45:41:30 \\
SA 20 SF3 & 00:00:37 & 45:52:07 \\
SA 20 SF4 & 00:00:35 & 46:06:39 \\
SA 20 SF5 & 00:00:56 & 45:52:35 \\
\hline \hline
\end{tabular}}
\label{tab:Landolt_obs}
\end{table}

Each night, calibration frames were taken along with science frames: flat, bias and dark frames. The color range of standard stars is chosen to be $B-V = (0.0, 1.3)$. In total, we have observed 31 Landolt fields. Due to the limitations imposed by the telescopes’s motors, the fields that were close to the zenith could not be observed. This resulted with observations of the Landolt’s fields only up to 60$^\circ$ height above the horizon.

\section{DATA REDUCTION}
\label{datareduction}
\indent

All raw images were first astrometrically solved and reduced. Astrometric solution was obtained using Astrometry software \citep{Lang2010}. Seven frames in the $B$-band were not successfully solved initially due to either insufficient flux or tracking errors. These frames were smoothed with Gaussian filter and then astrometrically solved. Data reduction was done following the standard procedure based on the Milankovic pipeline \citep{Muller2019}. Master bias frame was constructed for each night as a median of all bias frames and subtracted from each science frame. Master dark frame was created from individual dark frames of the largest exposure time for the particular night scaled to correspond to individual exposures. Master bias frame was subtracted from each flat field image, along with the master dark frame of the same exposure time (5 seconds exposure). Afterwards, each flat field image was normalized to its median value in each of the filters ($B, V, R$ and $I$) and the median stack was created as the final master flat. Then the science frames were divided by these master flat frames. These fully reduced science frames were then fed into the Python script that measured the magnitudes of the stars selected based on their celestial coordinates. Aperture photometry was done inside apertures \sout{2.2} 3.3 times larger then the FWHM = 4.5 pix = 2.7 arcsec using {\tt Photutils} package\footnote{\{{\tt Photutils} is an open source Python package (a part of the Astropy project) that provides tools including, but not limited to the aperture photometry (\url{https://photutils.readthedocs.io/en/stable/})}. In the sky annulus (10 pixels wide) around each star the mask was created to exclude any other object if present using sigma clipping. Instrumental magnitudes were measured according to the standard formula:
\begin{eqnarray}
\label{mag}
& m = -2.5 * \log (F) + 2.5  \log (t), \\
& F = \sum{C_i} - A s, 
\end{eqnarray}
where $C_i$ are counts measured in the aperture, $F$ is the flux, $A$ is the aperture area, $s$ is the sky measured inside sky annulus divided by its area (sky per pixel), and $t$ is the observational time in seconds.

Errors in magnitudes ($\Delta m$) were calculated using formula from IRAF's {\tt phot} package slightly modified:

\begin{eqnarray*}
&\Delta m = 1.0857  \Delta F / F, \\
&\Delta F = \sqrt{F + A  \sigma^2 + A^2 \sigma^2 / S},
\end{eqnarray*}
where $\sigma$ is the standard deviation of the background inside sky annulus and $S$ is the area of the sky annulus.
\begin{table*}
\caption{List of standard stars' fields selected for photometric calibration and observed at the ASV with 60 cm Nedeljkovi\'c telescope equipped with FLI PL 230 CCD camera. The first column is the Landolt's field; the second column lists a numuber of stars in that field; form the third one each three columns give, for a specific filter (f) indicated by the letter B, V, R or I: the air mass (X) and the exposure time of individual science frames in seconds (Exp). Since there are 4 different filters, there are 12 columns in total for each of the fields selected from the Landolt catalog of standard stars labeled in the first column.}
\vskip.25cm
\parbox{\textwidth}{
\centerline{\begin{tabular}{lcccccccccccccc}
\hline
Field & N & f & X & Exp & f & X & Exp & f & X & Exp & f & X & Exp\\
\hline
GD 2 & 6 & B & 1.01 & 180 & V &  1.01 & 100 & R &    1.01 & 100  & I & 1.01 & 100 \\
GD 8 & 4 & B & 1.08 & 600 & V &  1.09 & 300 & R &   1.07 & 300  & I &  1.06 & 300\\
GD 10 & 4 & B & 1.02 & 300 & V &  1.02 & 200 & R &  1.01 & 230  & I & 1.01 & 250 \\
GD 13 & 2 & B & 1.31 & 300 & V &   1.32 & 120 & R &  1.29 & 120  & I & 1.28 & 120\\
GD 275 & 2 & B & 1.45 & 300 & V &  1.56 & 200 & R &  1.43 & 200  & I & 1.40 & 200 \\
GD 277 & 3 & B & 1.39 & 300 & V &  1.35 & 180 & R  &  1.34 & 180  & I & 1.33 & 180\\
GD 278 & 3 & B & 1.20 & 300 & V &   1.21 & 180 & R  &  1.19 & 180  & I & 1.18 & 180\\
GD 279 & 10 & B & 1.11 & 90 & V &  1.11 & 45 & R &   1.10 & 45  & I & 1.10  & 45\\
GD 325 & 4 & B & 2.18 & 300 & V &    1.72 & 150 & R   &  1.78 & 150  & I &   2.15 &150\\
GD 336 & 4 & B & 1.52 & 120 & V &   1.54 & 60 & R   &  1.55 & 60  & I  &  1.57 & 60\\
GD 363 & 5 & B & 1.19 & 180 & V &  1.21 & 90 & R &    1.22 & 120  & I &  1.22 & 120\\
GD 378 & 4 & B & 1.20 & 130 & V & 1.22 & 80 & R   &  1.23 & 80  & I &  1.23 & 80\\
GD 391 & 9 & B & 1.06 & 230 & V &   1.07 & 100 & R   &  1.07 & 100  & I &   1.07 & 100 \\
GD 405 & 2 & B & 1.43 & 300 & V & 1.40 & 150 & R &  1.31 & 120  & I &  1.30 & 120\\
GD 421 & 5 & B & 1.18 & 180 & V &   1.20 & 120 & R   &  1.19 & 90  & I &  1.19 & 90\\
KUV 433 & 2 & B & 1.26 & 150 & V &   1.27 & 90 & R &   1.28 & 100  & I & 1.29  & 100\\
PG1430+427 & 4 & B & 1.51 & 120 & V  &  1.54 & 80 & R   &  1.58 & 30  & I  &  1.59 &30\\
PG1648+536 & 6 & B & 1.21 & 90 & V  &  1.22 & 25 & R &   1.23 & 10  & I &  1.23 &10\\
PG1343+578 & 2 & B & 1.92 & 500 & V &  1.89 & 200 & / &   /  & / & / & / & /\\
SA 20 SF1 & 4 & B & 1.02 & 30 & V   &  1.02 & 15 & R &    1.02 & 15  & I   &  1.02 & 15\\
SA 20 SF2 & 5 & B & 1.02 & 30 & V   &  1.02 & 15 & R   &  1.02 & 15  & I  &  1.02 & 15\\
SA 20 SF3 & 4 & B & 1.02 & 30 & V   &  1.02 & 15 & R   &  1.02 & 15  & I &  1.02 & 15\\
SA 20 SF4 & 5 & B & 1.02 & 50 & V   &  1.01 & 25 & R &   1.01 & 25  & I  &  1.01 & 25\\
SA 20 SF5 & 9 & B & 1.01 & 60 & V &   1.01 & 30 & R  &  1.01 & 30  & I  &  1.01 & 30 \\
SA 23 SF1 & 5 & B & 1.32 & 45 & V  &  1.31 & 20 & R   &  1.31 & 10  & I  &  1.30 & 5 \\
SA 23 SF2 & 5 & B & 1.30 & 45 & V   &  1.29 & 20 & R &    1.29 & 20  & I &  1.28 & 10\\
SA 35 SF1 & 4 & B & 1.29 & 10 & V   &  1.29 & 5 & R &    1.29 & 5  & I   &  1.30 & 5\\
SA 38 SF1 & 2 & B & 1.17 & 130 & V  &  1.18 & 60 & R  &  1.19 & 60  & I  &  1.19 & 60\\
SA 38 SF4 & 13 & B & 1.19 & 160 & V  &  1.21 & 20 & R &    1.22 & 10  & I  &  1.22 & 10 \\
SA 41 SF1 & 13 & B & 1.01 & 230 & V  &  1.02 & 200 & R &   1.02 & 200  & I  &  1.02 & 200 \\
SA 41 SF2 & 5 & B & 1.03 & 150 & V  &  1.04 & 20 & R &   1.04 & 20  & I &  1.04 & 20\\
\hline
\end{tabular}}}
\label{stdstars}
\end{table*}

In the Table \ref{stdstars}, all observed and processed  fields are listed. For each of the $BVRI$ filters the number of stars is indicated in the column labeled with N. PG1343+578 field was excluded from further analysis, since only $B$- and $V$-frames were imaged thus lacking color information that is needed for photometric calibration. In total, there are 154 stars in 30 Landolt fields for subsequent  photometric calibration. 

\section{PHOTOMETRIC CALIBRATION}
\label{calibration}
\indent

The aim of this work is to relate instrumental magnitudes measured at the ASV with 60 cm Nedeljkovi\'c telescope equipped with FLI PL 230 CCD camera to the magnitudes of the standard stars selected from the catalog of \cite{Landolt}. Calibrated magnitudes we wish to determine are those measured with the detector that perfectly matches the standard $BVRI$ system operating above the atmosphere. Both magnitudes and colors should match the chosen standard system measurements. Photometric calibration consists of correcting for two effects: (1) atmospheric extinction and (2) mismatch between instrumental and standard system. 

First, instrumental magnitudes need to be corrected for atmospheric extinction or dimming of the starlight by the earth's atmosphere. The longer the path of the light the more it is dimmed\footnote{A star closer to the horizon will dim more than the one closer to the zenith.}. The length of this path is called the air mass. We can relate the magnitude of the object above the atmosphere $m_0(\lambda)$ for any particular wavelength $\lambda$ to the one measured at the surface of the earth $m_i$:

\begin{equation}
\label{m0}
    m_0(\lambda) = m_i(\lambda) + k(\lambda) X(z),
\end{equation}
where $X(z)$ is the air mass and $k(\lambda)$ is the extinction coefficient at the zenith distance $z$. Air mass $X$ can be approximated for small zenith angles with $X=sec(z)$, where $z$ is the zenith distance. More precise relationship is given in \cite{Young}. Air mass is calculated automatically and added to the header of images during the observing session. The air mass range spans from 1 to 1.5. Extinction coefficient can be determined as the slope of the relation between the instrumental magnitude and the air mass. During the night, target stars will be at different air masses, and from the simple linear relation Eq.~(\ref{m0}) we can determine extinction coefficient for a specific filter used. Extinction coefficient is dependent on the wavelength since shorter wavelengths (blue light) scatter more then longer ones (red light); for each filter one needs to measure extinction coefficient separately. 

The second effect, a possible mismatch between our instrumentation and Landolt's, can be corrected by adding another term to the Eq.~\ref{m0}, the so-called color term ($\epsilon$):

\begin{equation}
\label{color_term}
    m_{\rm std}(\lambda) = m_0(\lambda) + \epsilon\ C,
\end{equation}
where $C$ is the color, and $m_{\rm std}$ the standard magnitude. This correction can be done only if more than one filter is being used, since color is the difference of magnitudes in two bands. For every combination of magnitudes and colors, one such equation can be written, thus a system of linear equations needs to be solved to standardize the instrumental system used.

Another problem is related the the blue band, since as we have already pointed out, shorter wavelengths scatter more than longer ones, and thus we might expect a dependence of the extinction coefficient on the color. In the blue band, another term can appear, the co-called secondary extinction coefficient ($k''$) that complicates equations further:
\begin{equation}
\label{sec_ext}
    k(\lambda) = k'(\lambda) + k''(\lambda)\ C.
\end{equation}

\subsection{Calibration without color correction}
\label{single}
We will consider here the case when the instrumental system is perfectly matched to the standard one. In this case, the only correction that needs to be applied is the extinction correction. This correction will give us magnitudes that are measured above the atmosphere with the combination of filters and a detector that match the one used in the standard $BVRI$ system.

Equations that relate instrumental magnitudes measured (Eq.~\ref{mag}) on the terrestrial surface to those that would be measured above the atmosphere, thus correcting for the dimming of starlight caused by light passage trough the atmosphere are:
\begin{eqnarray}
\label{air1}
b_0 = b_i - k_b X \\
v_0 = v_i - k_v X \\
r_0 = r_i - k_r X \\
i_0 = i_i - k_i X, 
\label{air}
\end{eqnarray}
where $BVRI$ denotes measured magnitudes (those with suffix 0 corresponds to the ones measured above atmosphere and suffix $i$ to instrumental magnitudes) and $k_{b,v,r,i}$ is the extinction coefficient in the corresponding band.

Matching them to the standard system reduces calibration equations to the set of the simple linear equations with only a constant offset (magnitude zero point) and a slope (extinction coefficient):
\begin{eqnarray}
\label{nocolor1}
B = b_0 + \zeta_b = b_i - k_b X + \zeta_b \\
V = v_0 + \zeta_v = v_i - k_v X + \zeta_v \\
R = r_0 + \zeta_r = r_i - k_r X + \zeta_r\\
I = i_0 + \zeta_i = i_i - k_i X + \zeta_i, 
\label{nocolor}
\end{eqnarray}
where $BVRI$ are the standard stars' magnitudes taken from the catalog, $\zeta$'s are the magnitude zero point offsets between the instrumental and the standard system, and $k$'s are the extinction coefficients in each $BVRI$ filter used. 

\subsection{Calibration with color correction}
\label{both}
Color shifts between ours ($BVRI$) and standard stars' filters ($BVRI$) can be quantified with the following set of equations:
\begin{eqnarray}
\label{color1}
B = b_i - k_b X + \epsilon_b (B-V) + \zeta_b \\
V = v_i - k_v X + \epsilon_v (B-V) + \zeta_v \\
R = r_i - k_r X + \epsilon_r (V-R) + \zeta_r\\
I = i_i - k_i X + \epsilon_i (V-I) + \zeta_i, 
\label{color}
\end{eqnarray}
where $\epsilon$'s are the color terms for particular colors. More equations of this kind can be written to address all possible combinations of filters used. Air mass $X$ is the average value of air masses corresponding to the air masses of observations taken with two different filters, since the fields cannot be imaged simultaneously. 

In the case of additional correction for extinction in the $B$-band, the extinction coefficient $k_b$ in the above equation should be replaced with the expression given by Eq.~(\ref{sec_ext}). Since this effect is very small, only in the case of well sampled data regarding  the color range, this correction should be applied.

\section{RESULTS}
\label{results}
\indent

All the equations derived for different corrections here, can be solved by multiple regression. However, since we have errors both in dependent and independent variables, we have chosen to apply orthogonal distance regression, that takes into account measurement errors in all the variables. 

We have used {\tt ODRPACK (ODR)} package from {\tt SciPy v1.6.1} library in Python \citep{scipy}. ODRPACK is actually a {\tt FORTRAN-77} library for performing orthogonal distance regression with possibly non-linear fitting functions \citep{boggs}. It is based on Levenberg-Marquardt-type algorithm including weights to account for different variances of the observations. 

\begin{figure*}[h]
    \centering
    \parbox{7.5cm}{
    \includegraphics[width=7.5cm]{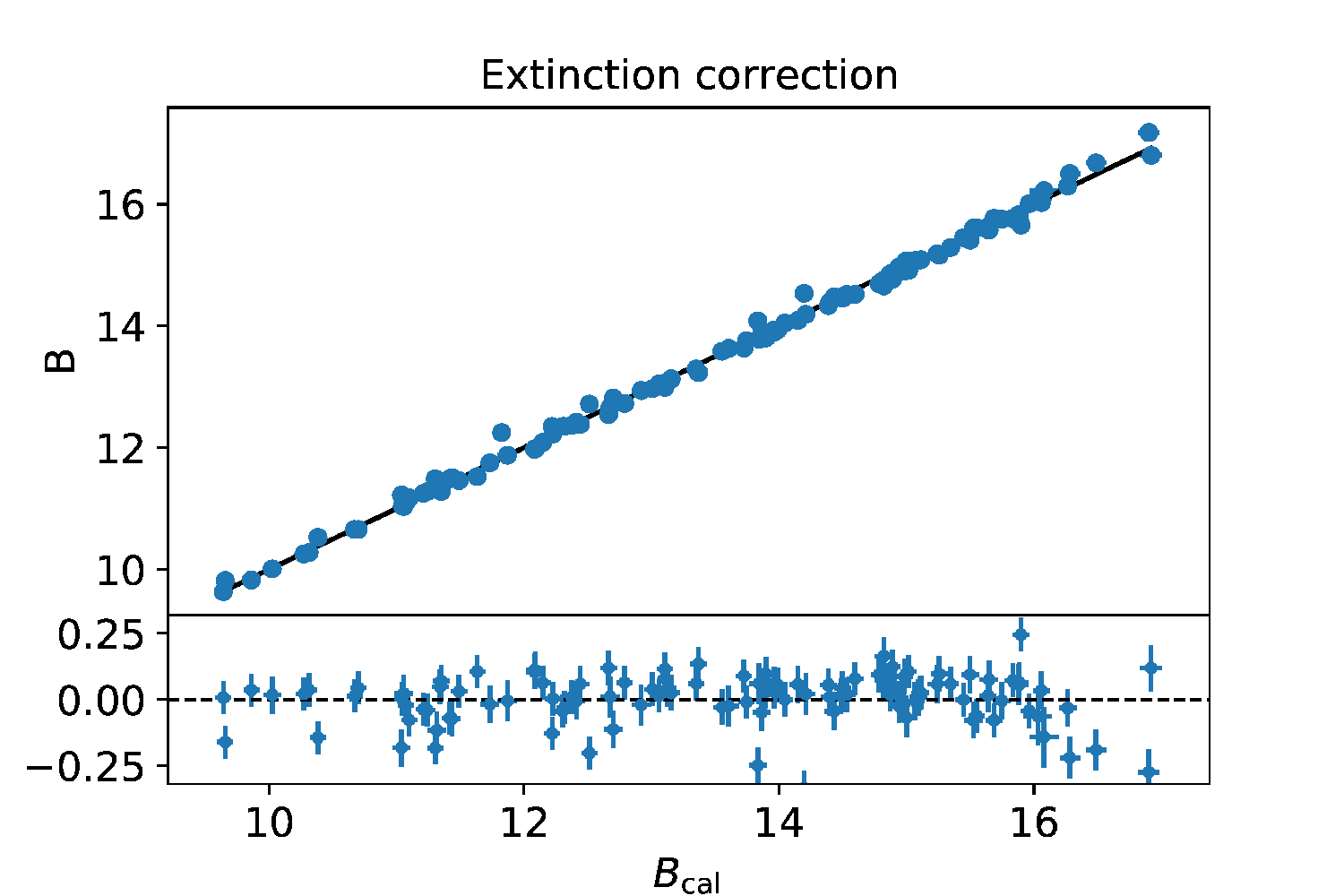}
    \includegraphics[width=7.5cm]{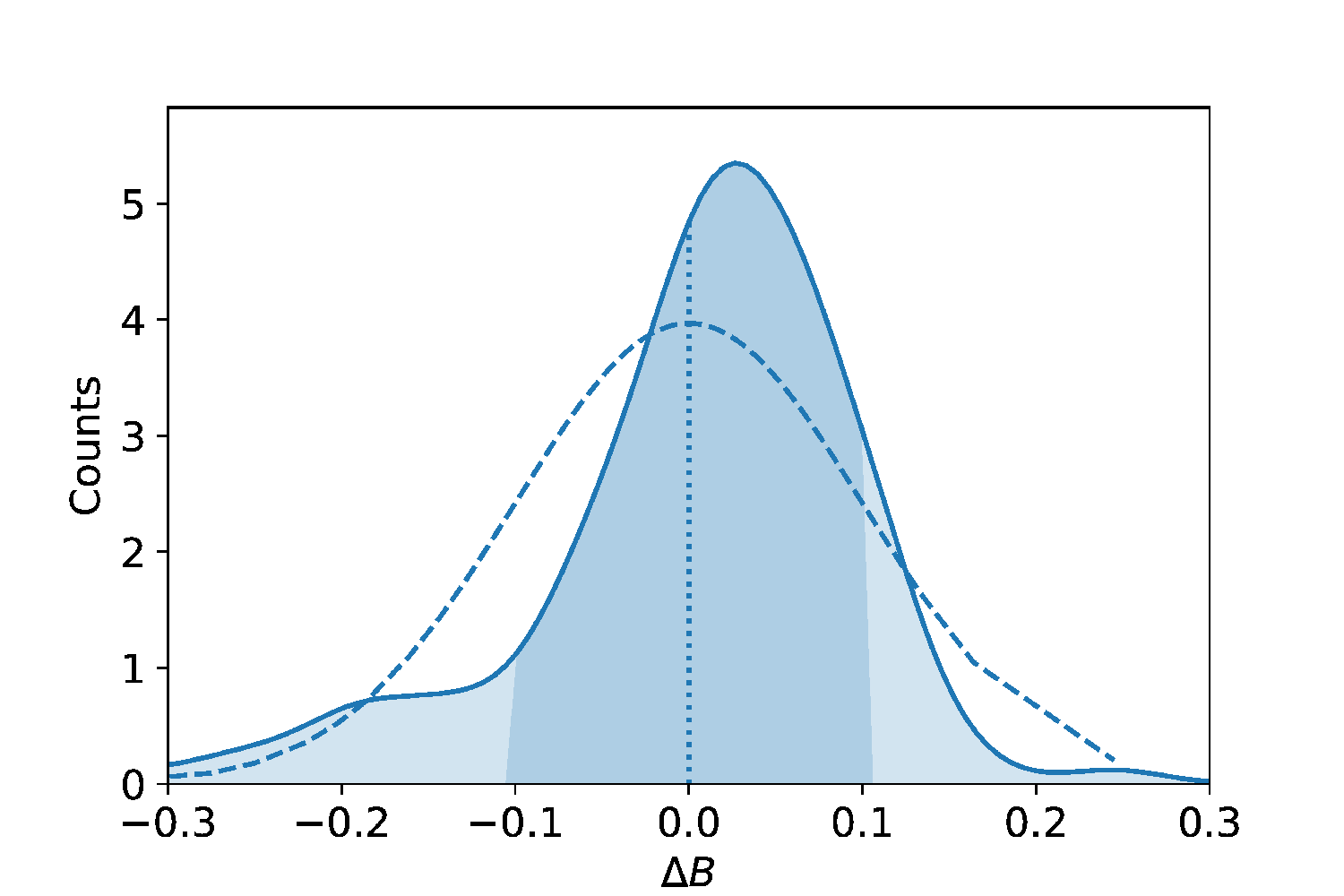}
\caption{Photometric calibration with only extinction correction: (up) calibrated $B$-band magnitude with residuals shown in the lower part of the plot, (down) residuals with shaded area corresponding to the standard deviation $\sigma=0.1$; dashed line is the normal distribution over-plotted.}
\label{fig:Bsimple}}
\qquad
\begin{minipage}{7.5cm}
    \includegraphics[width=7.5cm]{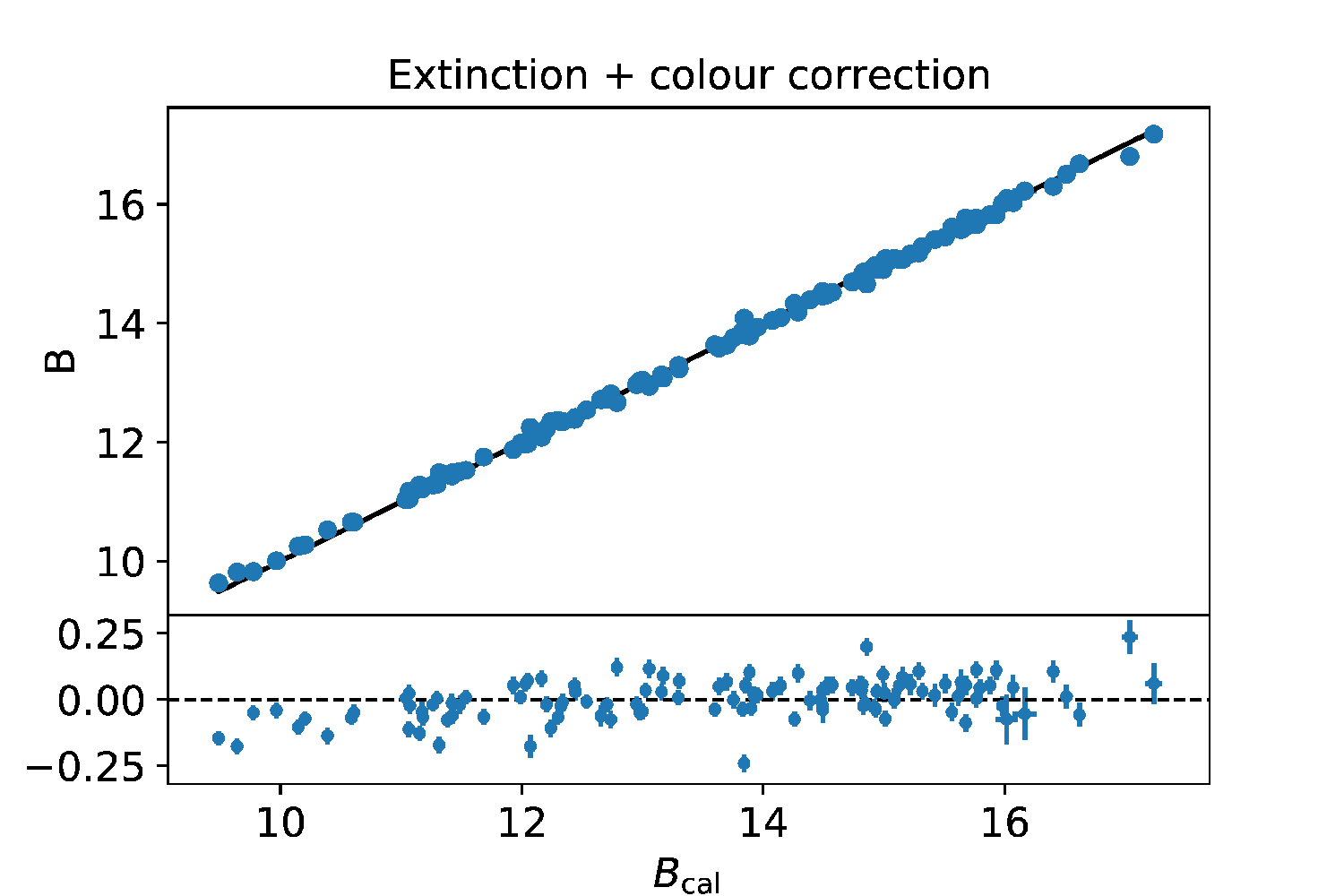}
    \includegraphics[width=7.5cm]{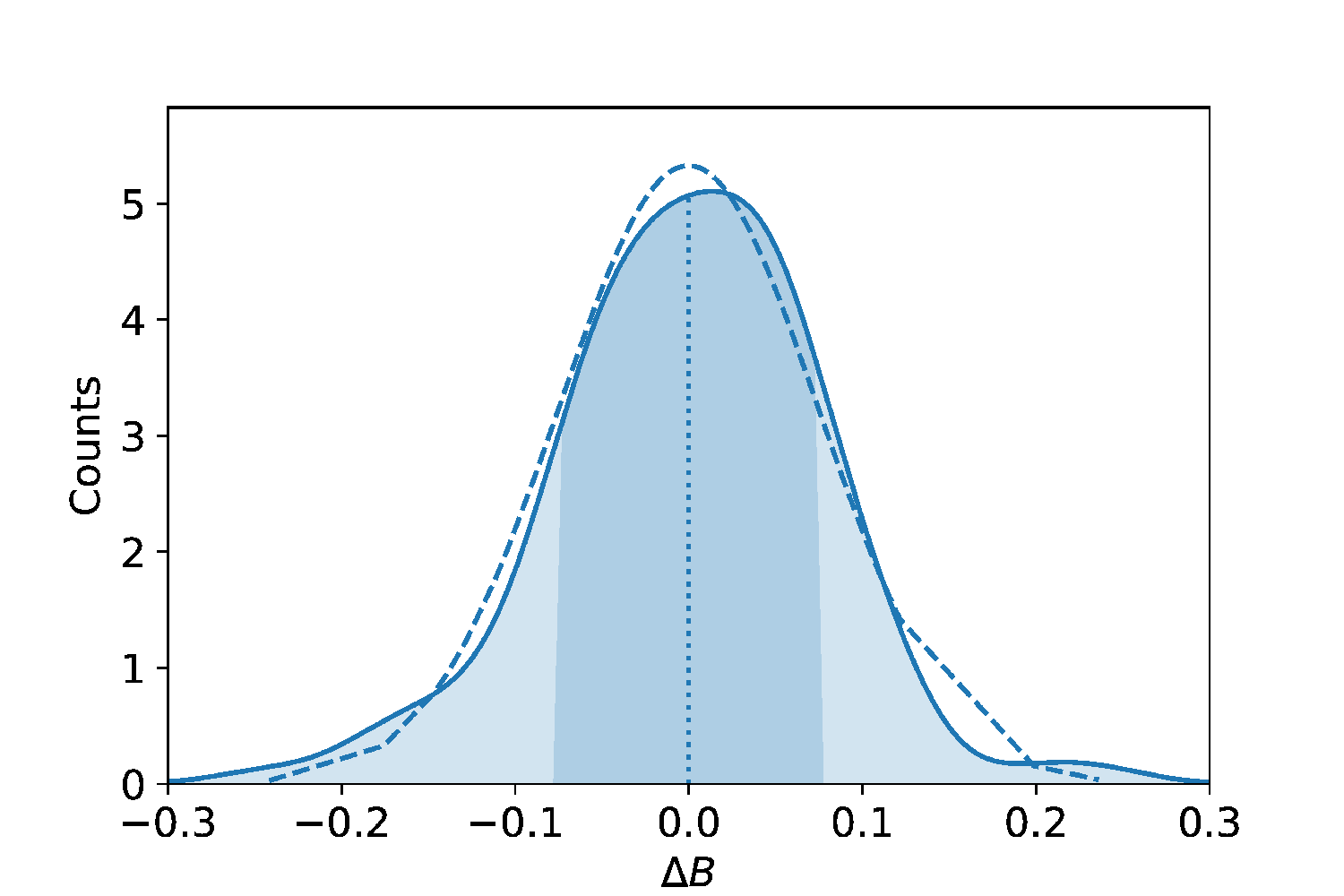}
    \caption{Photometric calibration with extinction and color correction: (up) calibrated $B$-band magnitude with residuals shown in the lower part of the plot, (down) residuals with shaded area corresponding to the standard deviation $\sigma=0.08$; dashed line is the normal distribution overplotted.}
    \label{fig:Bmulti}
\end{minipage}
\end{figure*}

Photometric calibration with and without the color term was done by creating a model and applying an orthogonal distance regression to it. The code is available at Github.\footnote{\url{https://github.com/anavudragovic/photometry/stdphot_secext.py}} In the simple case of correcting only for the extinction (subsection \ref{single}), calibrated magnitude in the $B$-band, along with residuals $\Delta B = B_{\rm cal} - B$ (where $B_{\rm cal}$ stands for calibrated magnitude and $B$ refers to the standard stars magnitude) is given in the Fig.~{\ref{fig:Bsimple}}. In the more complex case with the color correction (subsection \ref{both}), calibrated magnitude with distribution of the residuals is given in the Fig.~{\ref{fig:Bmulti}}. Residuals are always the difference between the calibrated magnitude and the standards' stars one; only the calibrated magnitude will consist of simply the magnitude zero point and the extinction coefficient in the simple case of extinction correction, while it will have in addition the color term correcting for the color dependence in the more complicated case with color correction that we apply and discuss its importance.

In the $B$-band, it can be easily seen comparing residuals in the Fig.~\ref{fig:Bsimple} and Fig.~\ref{fig:Bmulti} that the error of the calibrated magnitude is smaller and that distribution of residuals is closer to normal distribution taking into account the color correction. The errors consist of measurement errors and statistical errors from the fitted parameters added in the quadrature. So, the reason the errors in the latter case are smaller simply refers to the smaller errors of the fitted parameters (magnitude zero point and extinction coefficient).  

\begin{figure*}[h]
    \centering
    \parbox{7.5cm}{
    \includegraphics[width=7.5cm]{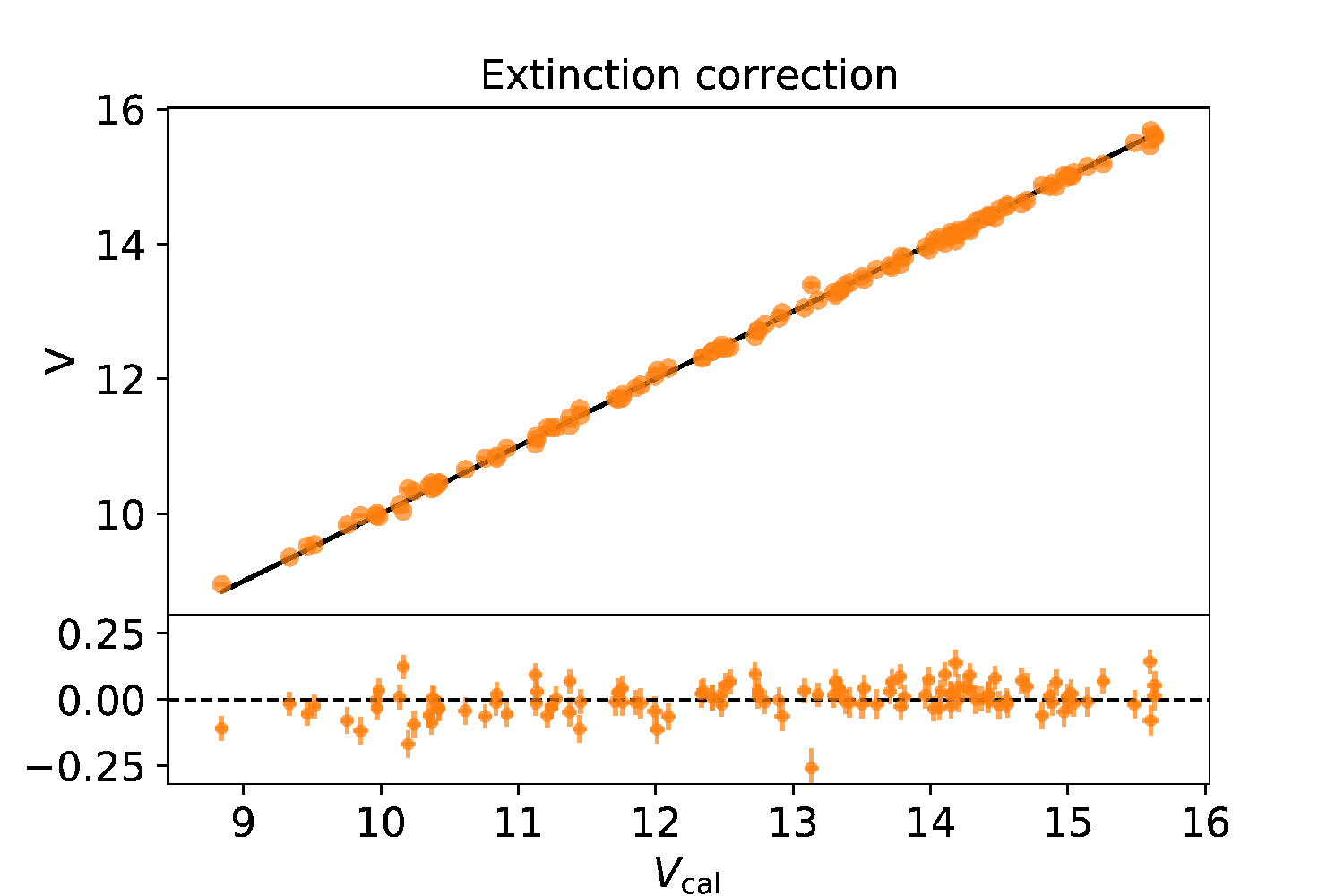}
    \includegraphics[width=7.5cm]{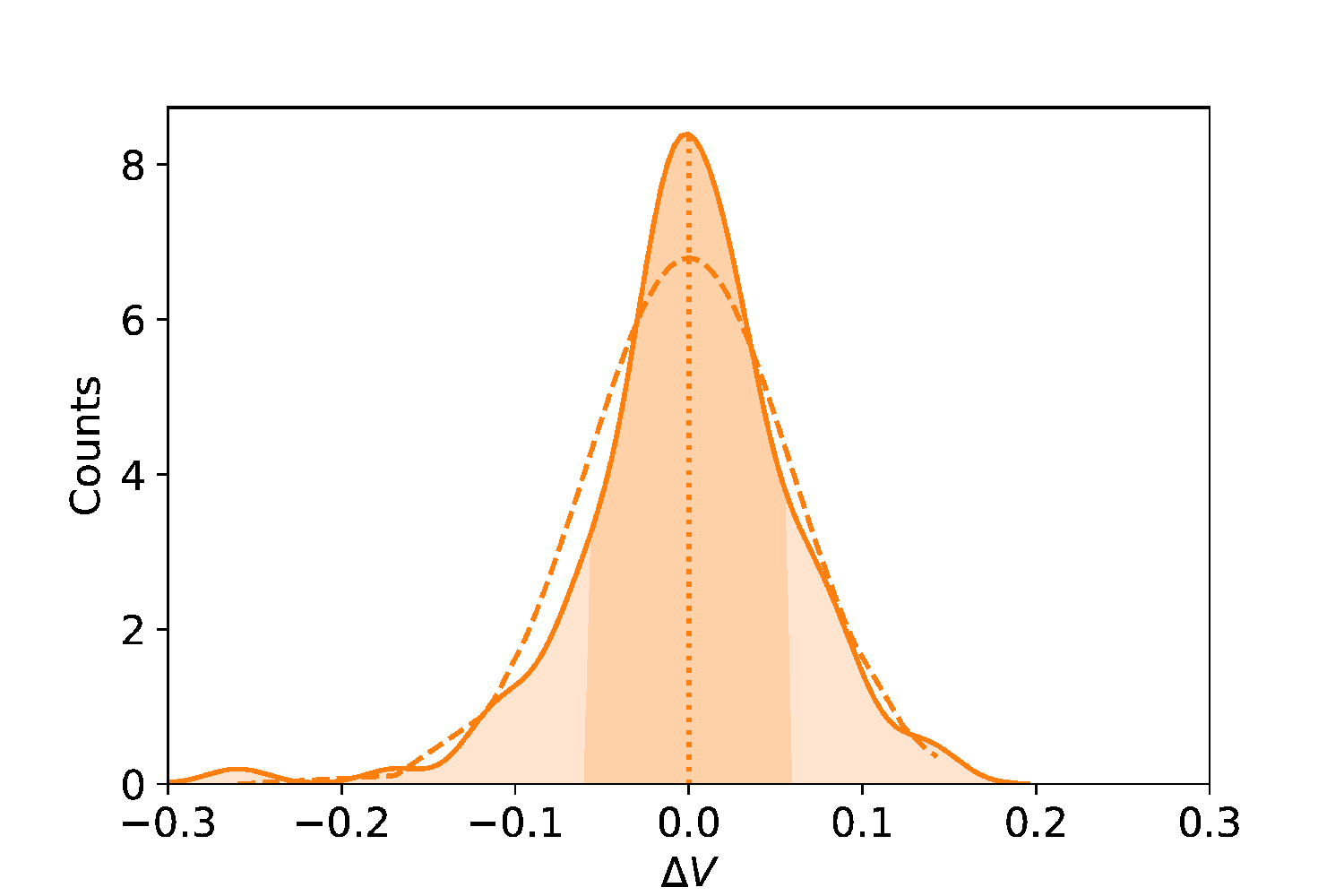}
\caption{Photometric calibration with only extinction correction: (up) calibrated $V$-band magnitude with residuals shown in the lower part of the plot, (down) residuals with shaded area corresponding to the standard deviation $\sigma=0.06$; dashed line is the normal distribution over-plotted.}
\label{fig:Vsimple}}
\qquad
\begin{minipage}{7.5cm}
    \includegraphics[width=7.5cm]{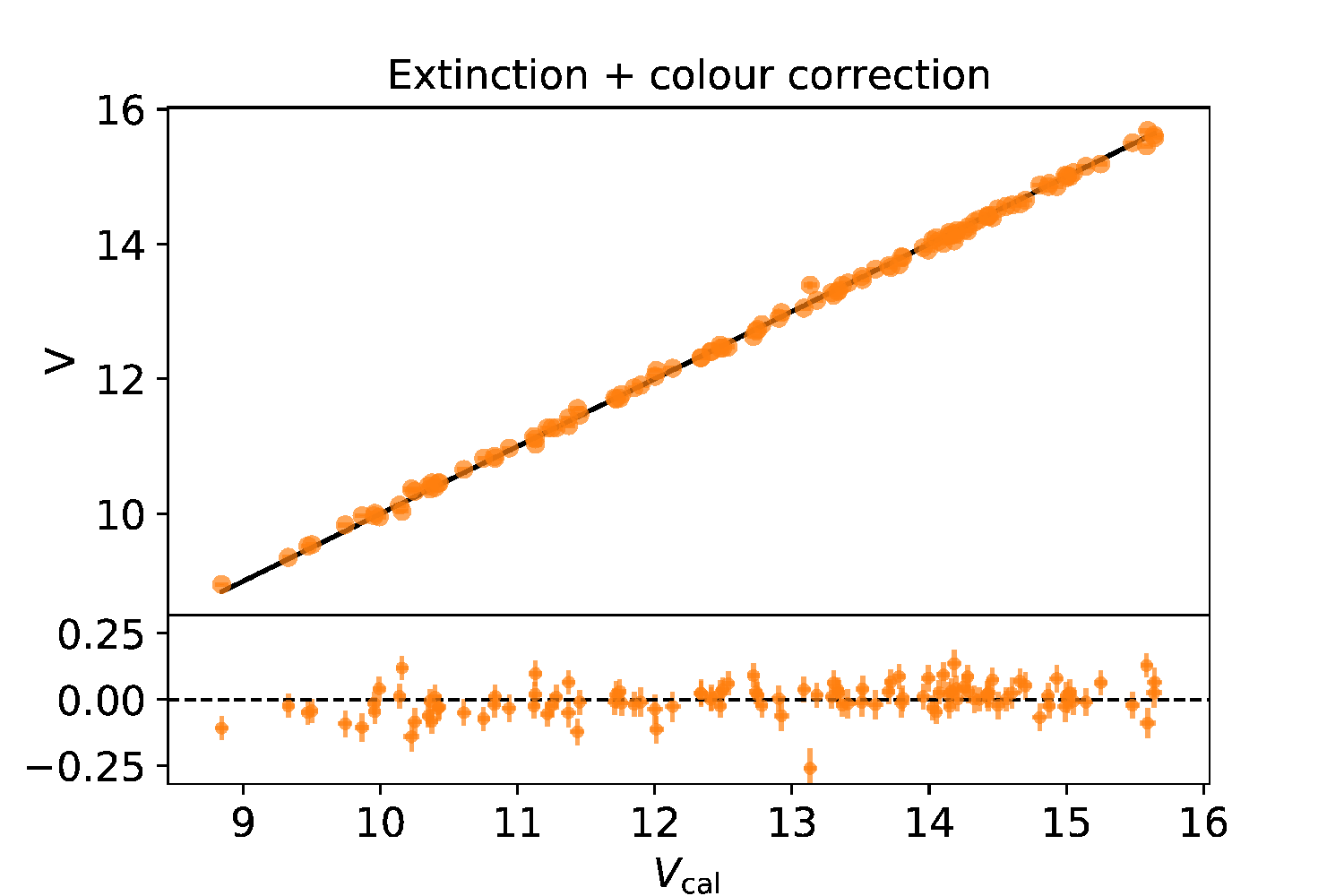}
    \includegraphics[width=7.5cm]{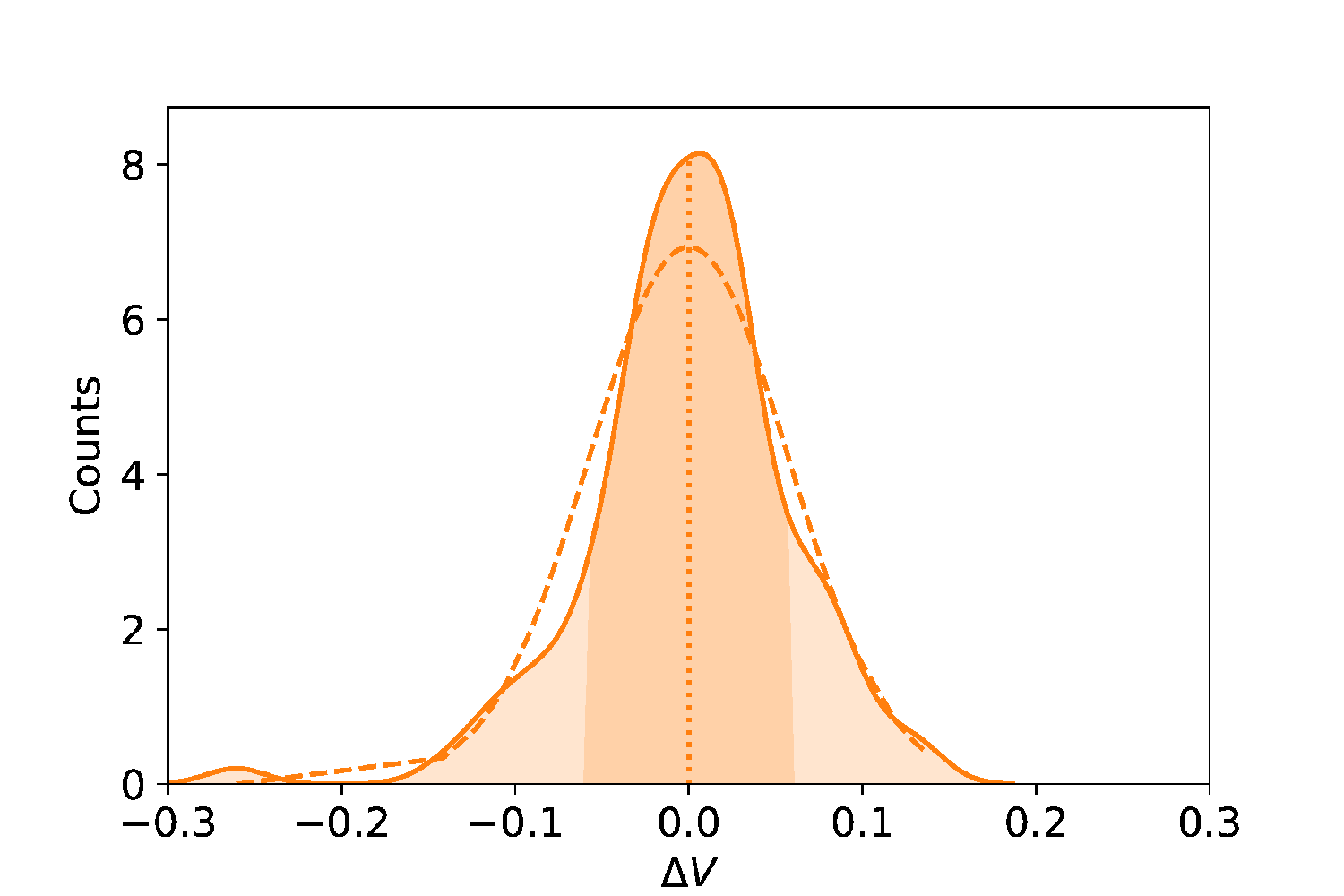}
    \caption{Photometric calibration with extinction and color correction: (up) calibrated $V$-band magnitude with residuals shown in the lower part of the plot, (down) residuals with shaded area corresponding to the standard deviation $\sigma=0.06$; dashed line is the normal distribution over-plotted.}
    \label{fig:Vmulti}
\end{minipage}
\end{figure*}

To test the significance of the level at which the two residual distributions (with and without the color correction) are close to the normal one, we have performed three statistical test: Shapiro-Wilk (\cite{shapiro}; {\tt shapiro} SciPy function), Anderson-Darling (\cite{adtest}; {\tt anderson} SciPy function) and D'Agostino's K$^2$ test (\cite{agostino}; {\tt normaltest} SciPy function). These are most commonly used statistical tests, each of which takes different assumptions and considers different aspect of data. They all test null hypothesis -- the normality of the residual distribution. Each test delivers both statistics and a p-value; statistics is compared to some pre-calculated critical value for the particular test, and p-value is a measure of the probability that the difference between distribution of residuals and normal distribution (in our case) may occur by a random chance: higher the p-value, stronger is the evidence in favor of the null hypothesis and vice versa. In the SciPy implementation of these tests, interpretation is the following: if p $>\alpha$ than our assumption holds. The significance value $\alpha$ to which we compare the measured p-value is some (predefined) probability of rejecting the null hypothesis that is actually true; a probability of making a wrong decision. This parameter is by convention set to $\alpha=0.05$, meaning that there is a probability of 5\% that we claim the residual distribution is normal, while it is not. 

To summarize, we will measure p-values for each of the statistical tests and compare it to the $\alpha$ value. If the p-value is larger than the $\alpha$ value, than our null hypothesis (assumption that residuals follow normal distribution) holds. Why is this important? Residuals should be randomly distributed, or else either the model is not correct (calibration equations) or our data sample is small or simply data are not well distributed over the range of our dependent variable(s). The consequence is that our model cannot successfully predict (calibrate) magnitudes given the particular data set. Normality assumption, along with calculated errors can also help us choose the most appropriate model for our data. On the other hand, errors of the fitted parameters should also be considered, since in the case they are comparable to the parameters themselves, models that introduce them should be disfavored compared to  simpler ones. We need to take all this into account when choosing the model that is the most appropriate for the given data set.

\begin{figure*}[h]
    \centering
    \parbox{7.5cm}{
    \includegraphics[width=7.5cm]{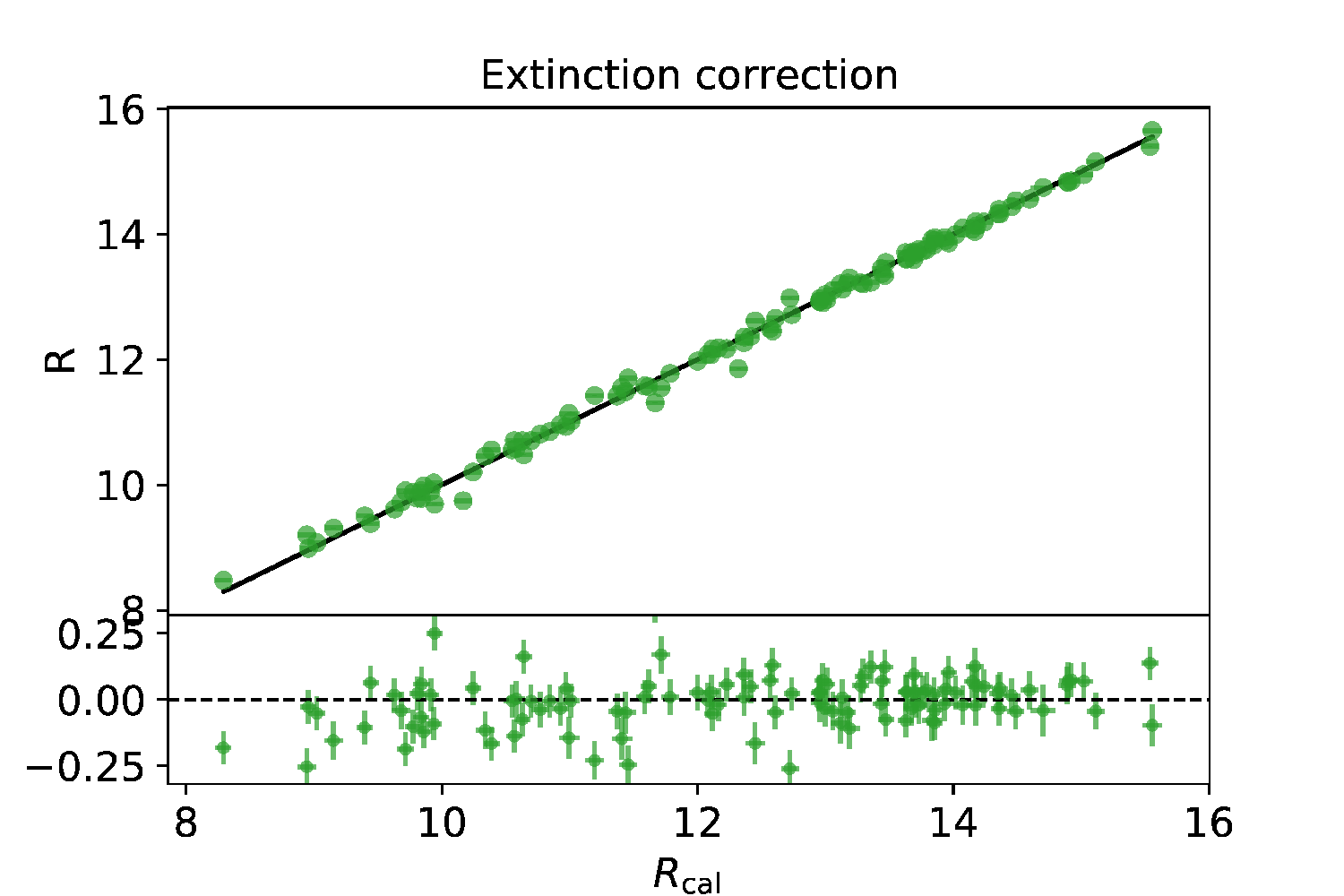}
    \includegraphics[width=7.5cm]{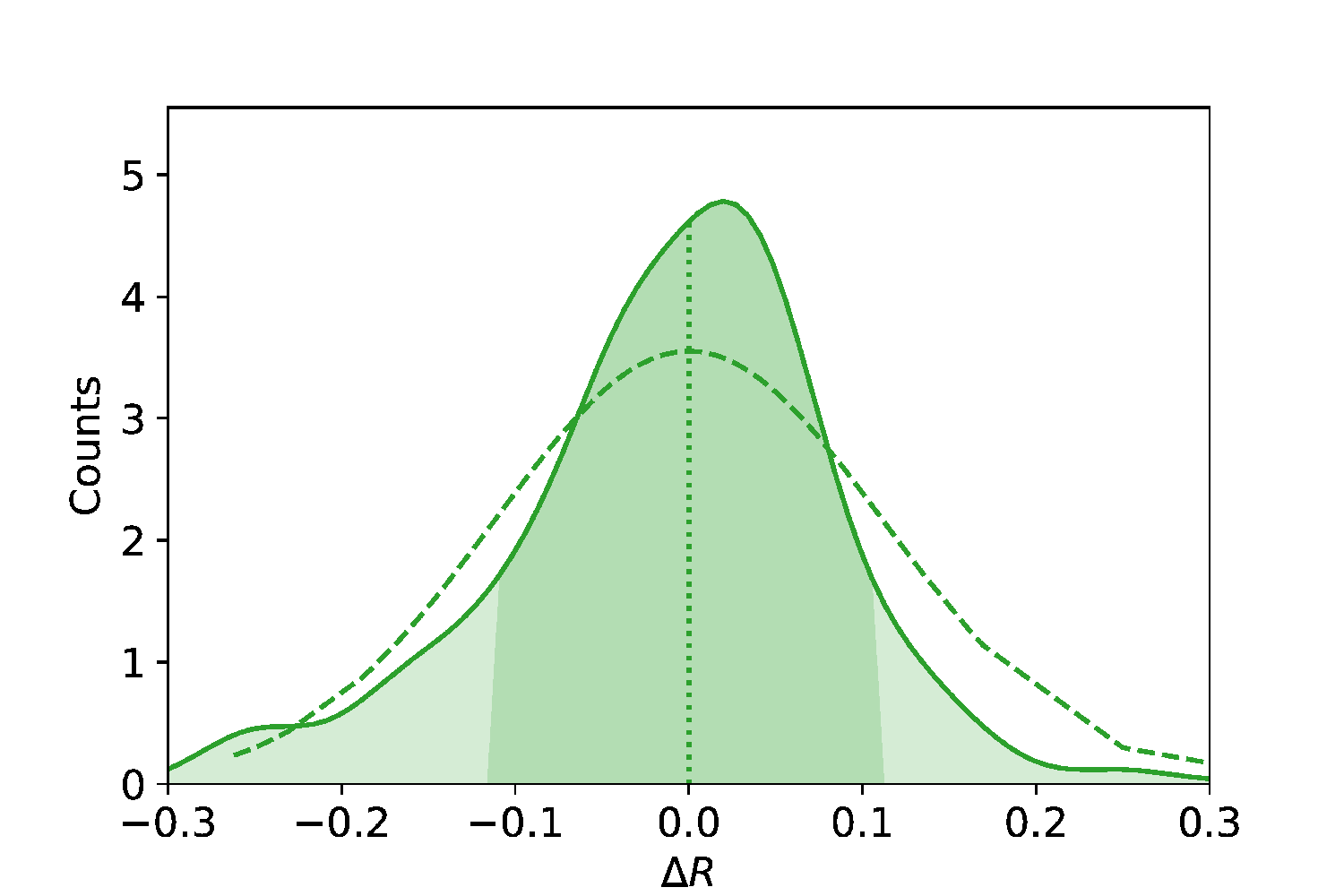}
\caption{Photometric calibration with only extinction correction: (up) calibrated $R$-band magnitude with residuals shown in the lower part of the plot, (down) residuals with shaded area corresponding to the standard deviation $\sigma=0.1$; dashed line is the normal distribution over-plotted.}
\label{fig:Rsimple}}
\qquad
\begin{minipage}{7.5cm}
    \includegraphics[width=7.5cm]{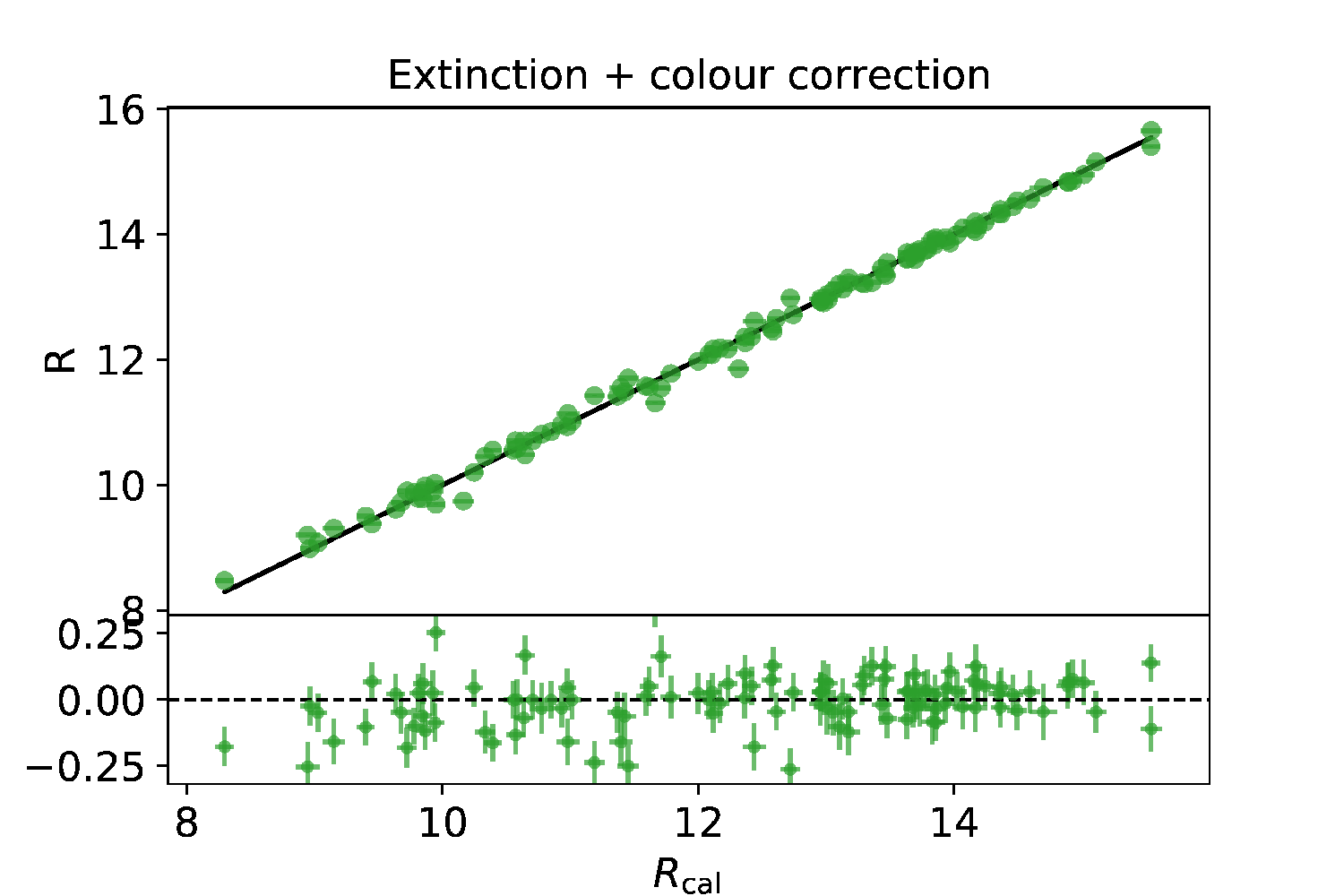}
    \includegraphics[width=7.5cm]{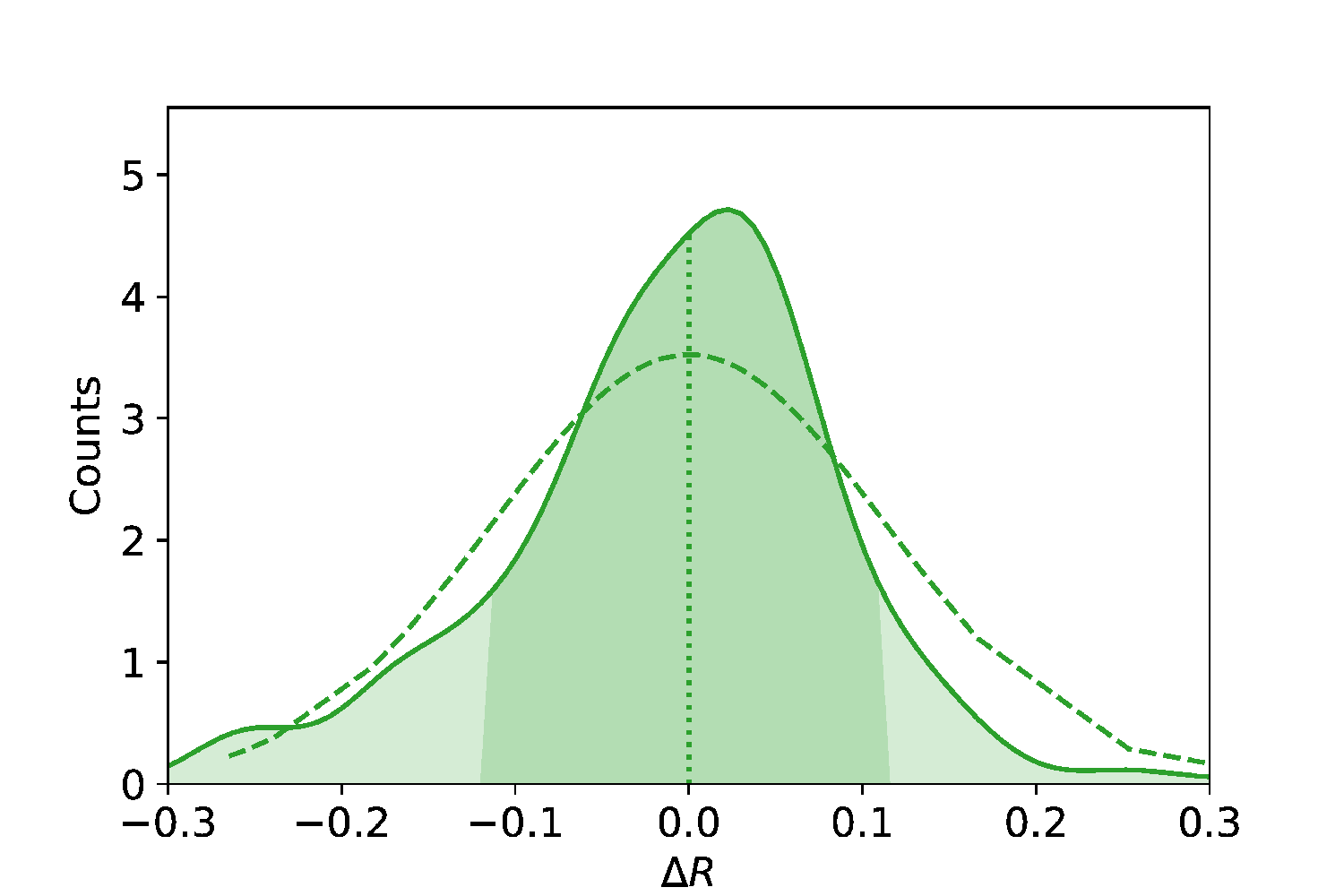}
    \caption{Photometric calibration with extinction and color correction: (up) calibrated $R$-band magnitude with residuals shown in the lower part of the plot, (down) residuals with shaded area corresponding to the standard deviation $\sigma=0.1$; dashed line is the normal distribution over-plotted.}
    \label{fig:Rmulti}
\end{minipage}
\end{figure*}

We run all three tests, since they inspect different features of the residual distribution they are fed with. Anderson-Darling test, for example, pays more attention to the wings of the distribution, and these are the areas where we visually see the difference from the normal distribution (in all the figures with residuals in all the bands, left wing is somewhat skewed). We shall demand that all the tests agree to accept the normality assumption for the distribution of the residuals.

\begin{figure*}[h]
    \centering
    \parbox{7.5cm}{
    \includegraphics[width=7.5cm]{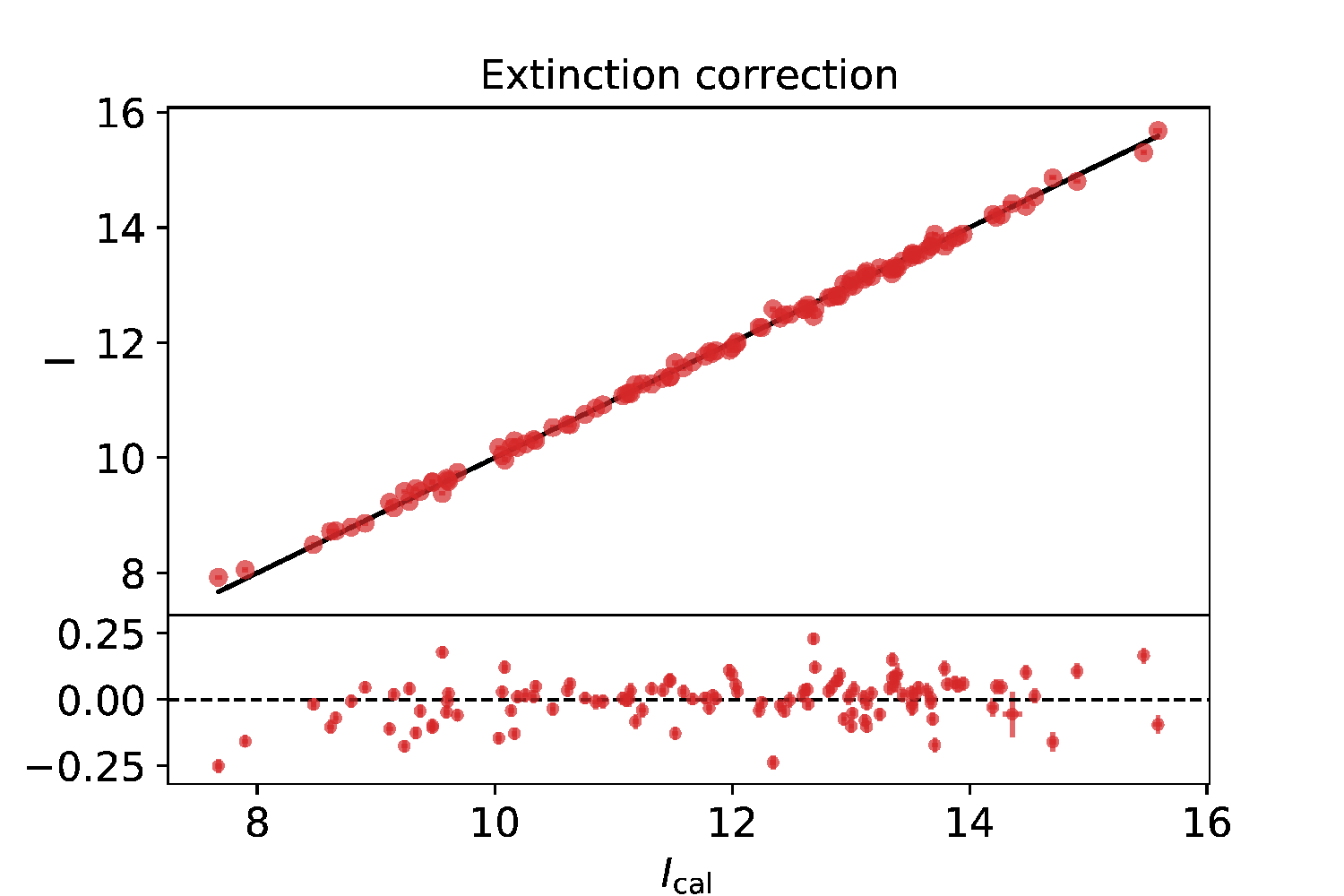}
    \includegraphics[width=7.5cm]{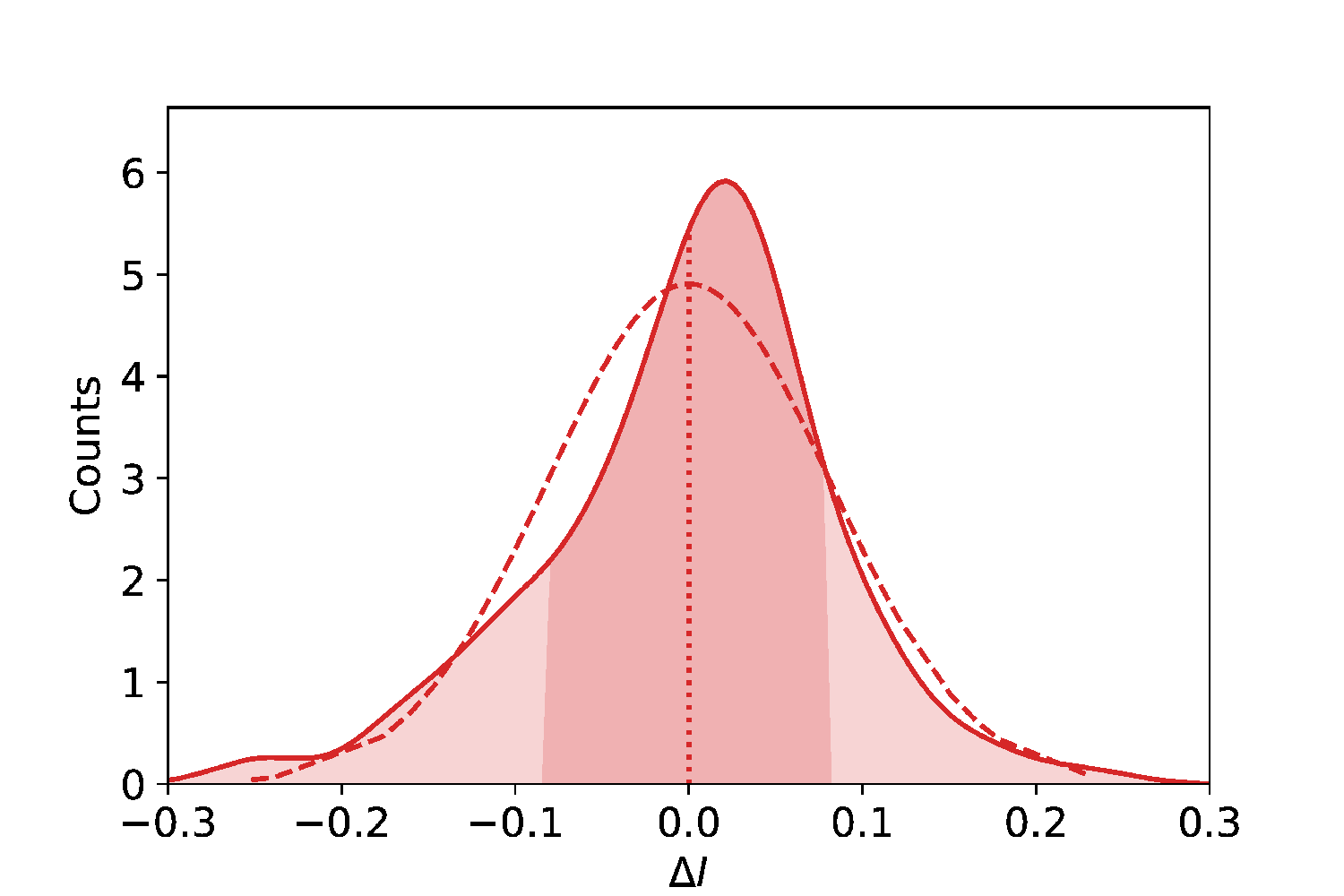}
\caption{Photometric calibration with only extinction correction: (up) calibrated $I$-band magnitude with residuals shown in the lower part of the plot, (down) residuals with shaded area corresponding to the standard deviation $\sigma=0.08$; dashed line is the normal distribution over-plotted.}
\label{fig:Isimple}}
\qquad
\begin{minipage}{7.5cm}
    \includegraphics[width=7.5cm]{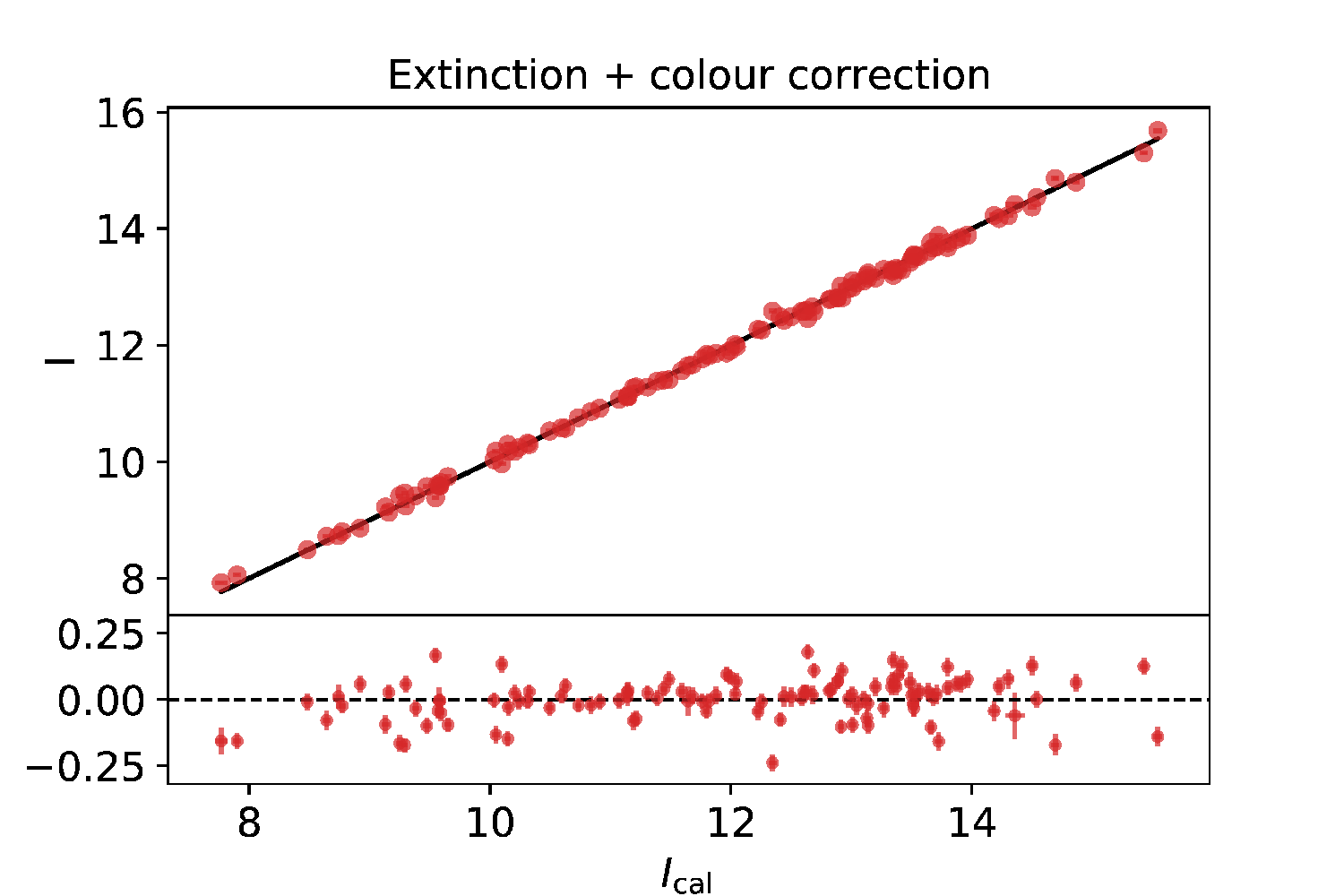}
    \includegraphics[width=7.5cm]{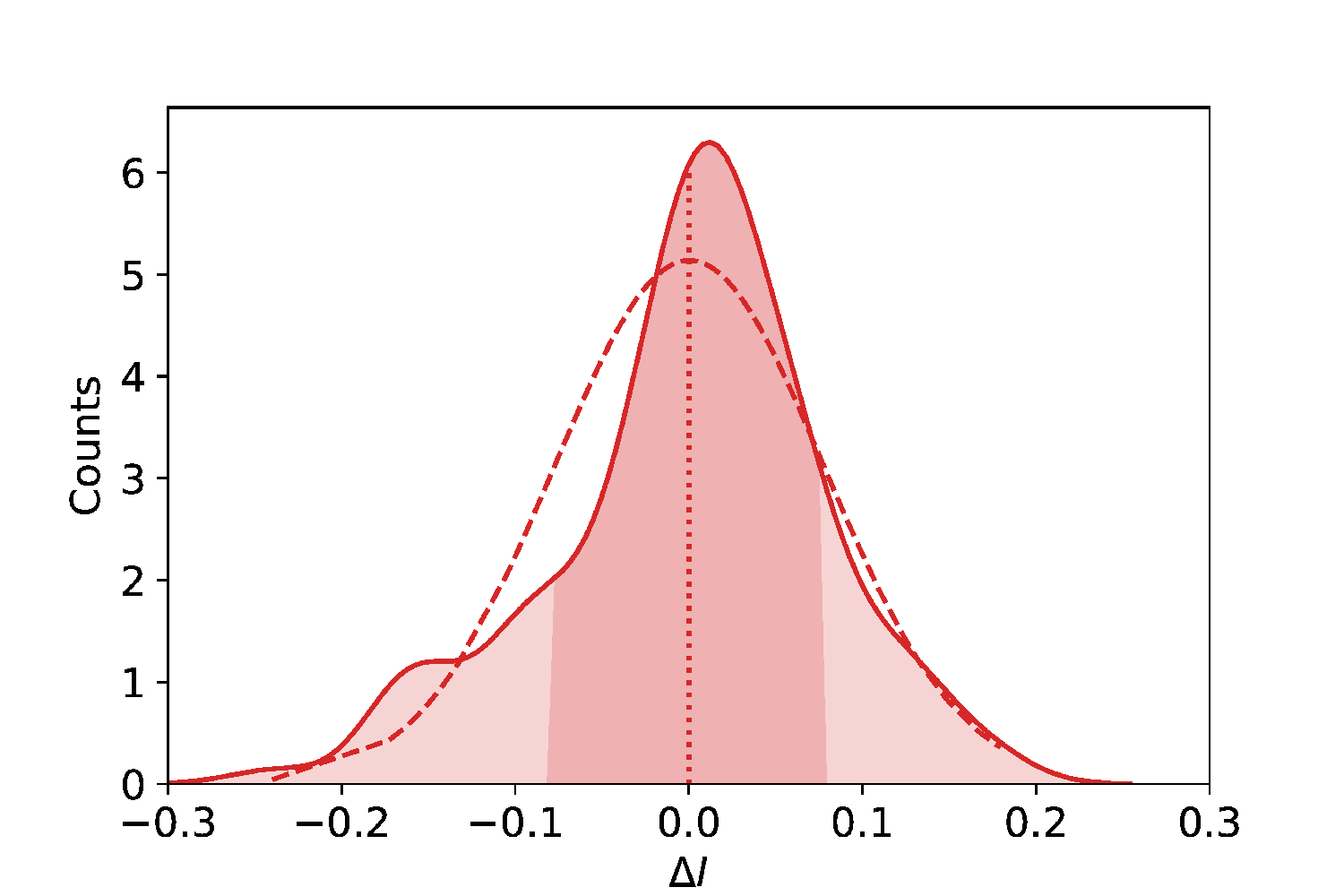}
    \caption{Photometric calibration with extinction and color correction: (up) calibrated $I$-band magnitude with residuals shown in the lower part of the plot, (down) residuals with shaded area corresponding to the standard deviation $\sigma=0.08$; dashed line is the normal distribution overplotted.}
    \label{fig:Imulti}
\end{minipage}
\end{figure*}

All the tests reject the normality assumption for the simple extinction correction in the $B$-band (Fig.~\ref{fig:Bsimple}), and fail to reject the same assumption with color correction applied (Fig.~\ref{fig:Bmulti}) at the significance level of $\alpha$=0.05. For the $V$-band, there is no difference between the tests performed on the residuals in both cases (with and without color correction). In the $R$- and $I$-band distribution of residuals in not normal in either case: with or without the color correction, so there is no need to apply the additional correction either (Fig.~\ref{fig:Rsimple} - \ref{fig:Imulti}). In addition, the standard deviation of the residuals is the same at 0.01 magnitude level (1\% accuracy) for the $V$-, $R$- and $I$-band.

\begin{table}
\caption{Results of the photometric calibration: $\zeta$ is the magnitude zero point for a given filter $f$=BVRI in the first column, $k$ is the extinction coefficient in the second column and $\epsilon$ is the color term given in the last column. Those rows that are given in bold are the most appropriate for given filters.}
\vspace{.2cm}
\centerline{\begin{tabular}{|c|c|c|}
\hline
$\zeta_{\rm f}$\small{[mag]} & $k_{\rm f}$\small{[mag/air mass]} & $\epsilon_{\rm f}$\\ 
\hline 
\multicolumn{3}{c}{B} \\ \hline 
21.78 $\pm$ 0.05 &  0.48 $\pm$ 0.04 & / \\ \hline
\textbf{21.32 $\pm$ 0.02} &  \textbf{0.19 $\pm$ 0.01} &  \textbf{0.17 $\pm$ 0.02} \\ \hline
\multicolumn{3}{c}{V} \\ \hline 
\textbf{22.11 $\pm$ 0.03} &  \textbf{0.37 $\pm$ 0.03} & \textbf{/}  \\ \hline 
22.08 $\pm$ 0.03 &  0.36 $\pm$ 0.03 &  0.02 $\pm$ 0.01   \\ \hline
\multicolumn{3}{c}{R} \\ \hline 
\textbf{22.12 $\pm$ 0.05} &  \textbf{0.48 $\pm$ 0.04} & \textbf{/}  \\ 
\hline 
22.15 $\pm$ 0.05 &  0.51 $\pm$ 0.04 &  0.01 $\pm$ 0.04 \\ \hline
\multicolumn{3}{c}{I} \\ \hline 
\textbf{21.26 $\pm$ 0.02} &  \textbf{0.16 $\pm$ 0.01} & \textbf{/}  \\ \hline
21.20 $\pm$ 0.02 &  0.15 $\pm$ 0.01 &  0.06 $\pm$ 0.02 \\ \hline
\end{tabular}}
\label{tab:coeffs}
\end{table}

We have also tested additional correction, the second order extinction correction in the $B$-band (Eq.~\ref{sec_ext}). Although the residuals behave well and the normality assumption is satisfied, the errors of the fitted parameters grew large, in some cases even larger than parameters themselves, implying that this additional effect cannot be measured from these data. Maybe if the data spanned a larger range in colors, additional correction could be applied. 

We conclude that the color correction is relevant only for the $B$-band. In the $V$-, $R$- and $I$-band simple extinction correction suffices. Magnitude zero points, extinction coefficients, and a single color term (in the $B$-band) with their corresponding errors are given in the Table \ref{tab:coeffs}. For the purpose of comparison, we give measured parameters in both cases, with and without the color correction, but we made bold those that are correct considering both residual distribution and errors of the measured parameters. For example, in the $B$-band, all errors of the fitted parameters get smaller when the color correction is applied (the first two rows in the Table \ref{tab:coeffs}). It can be also seen visually in the bottom of the upper panel in the Fig.~\ref{fig:Bmulti} compared to the Fig.~\ref{fig:Bsimple}. Moreover, lower panels of the same two figures representing distribution of the residuals of the calibrated vs.~true (standard stars) magnitude reveal different behaviour of the two models (without and with the color correction) with respect to the normal distribution (dashed line), reflected in the different p-value: 0.00 vs.~0.01 for the Shapiro-Wilk test or 0.00 vs.~0.093 for D'Agostino's K$^2$ test. These p-values should be larger than $\alpha$=0.05 and this is true in the case when color correction is applied for both tests. For the Anderson-Darling test, the critical value of the test at the 5\% significance level (0.76) is compared to the measured one in both cases (without and with the color correction): 3.355 $>$ 0.76 vs.~0.537 $<$ 0.76; in the first case, the measured value being larger than the critical one implies that the distribution of residuals is not normal, while in the latter case it implies it is. All the tests agree upon the normality assumption in the latter case and this model, the one with the color correction is favored over the more simpler one and given in bold in the Table~\ref{tab:coeffs}. 
In all other bands ($V,R,I$) the statistics doesn't play a big role. Both residual distributions (without and with color correction) are not normal according to all statistical tests. There is additional argument why simpler model is more appropriate for these bands. In the Table~\ref{tab:coeffs} for all except for the $B$-band, the additional color term (normal font line, not bold) has the error that is comparable or even larger than the value of the parameter, which means it cannot be determined from these data. So, for all other bands, we favour simple extinction correction (bold in the Table~\ref{tab:coeffs}). 

We suspect the reason for unsuccessful determination of color terms is related to the modest range in colors $B-V=(0.0,1.3)$. We lack both genuinely blue ($B-V<0$) and red stars ($B-V>1.4$) in our data set. On the other hand, this color range corresponds to the color of RU Camelopardalis (RU Cam) star that was the primary motivation for accurate photometric calibration, so our results can be applied to this star.

RU Cam is a long period puslating star, a W Virginis subtype of Type II Cepheid. In the past it has shown changes in the amplitude of the pulsation and length of the pulsation cycle on a scale that is decades long. We are observing the light curve in different filters to determine the nature of these changes. RU Cam is just one of the Type II Cepheids and anomalous Cepheids that are going to be observed at the ASV. Some Type II Cepheids show small changes in their amplitude that repeat every second period, so the phenomena is called period doubling (PD). The detection of requires a precision in the order of magnitude of a few hundredth of magnitudes. To study these changes it is required that we observe these stars for a very long time -- decades or longer. Long observations like that mean that the telescopes that observe the stars, as well as the observing equipment, will change, and the easiest way to overcome the different compensation among different systems is to anchor each and every observation to a standard photomeric system. The big automated observing programs (for example Optical Gravitational Lensing Experiment\footnote{\url{http://ogle.astrouw.edu.pl/}} use  either the $BVRI$ filters or the $ugriz$ filters. The standard photometry also becomes important when one takes into account that these pulsating stars change their effective temperature during the pulsation cycle. This would mean that finding an appropriate comparison star (if there would be one in the same observing field) would only approximate the spectral type of the pulsating star. Using the standard photometric calibration we deal with the color correction, increase the precision of our photometry, make it accessible for other long term programs to incorporate, and in the end it is also the long used consensus in the variable star community.

\section{PHOTOMETRIC CONDITIONS}
Photometric calibration due to high precision demands clear, dark skies. There are no previous measurements of the sky darkness at ASV. We have measured the surface brightness of the sky during observations as a mean flux of the median values inside dozen boxes 50 times 50 pixels large in the areas without objects. These values are then converted into the surface brightness by dividing the flux with the area of one pixel in arcseconds and then calibrated using photometric calibration coefficients from the Table~\ref{tab:coeffs}.

\begin{figure}[h!]
    \centering
    \includegraphics[width=.5\textwidth]{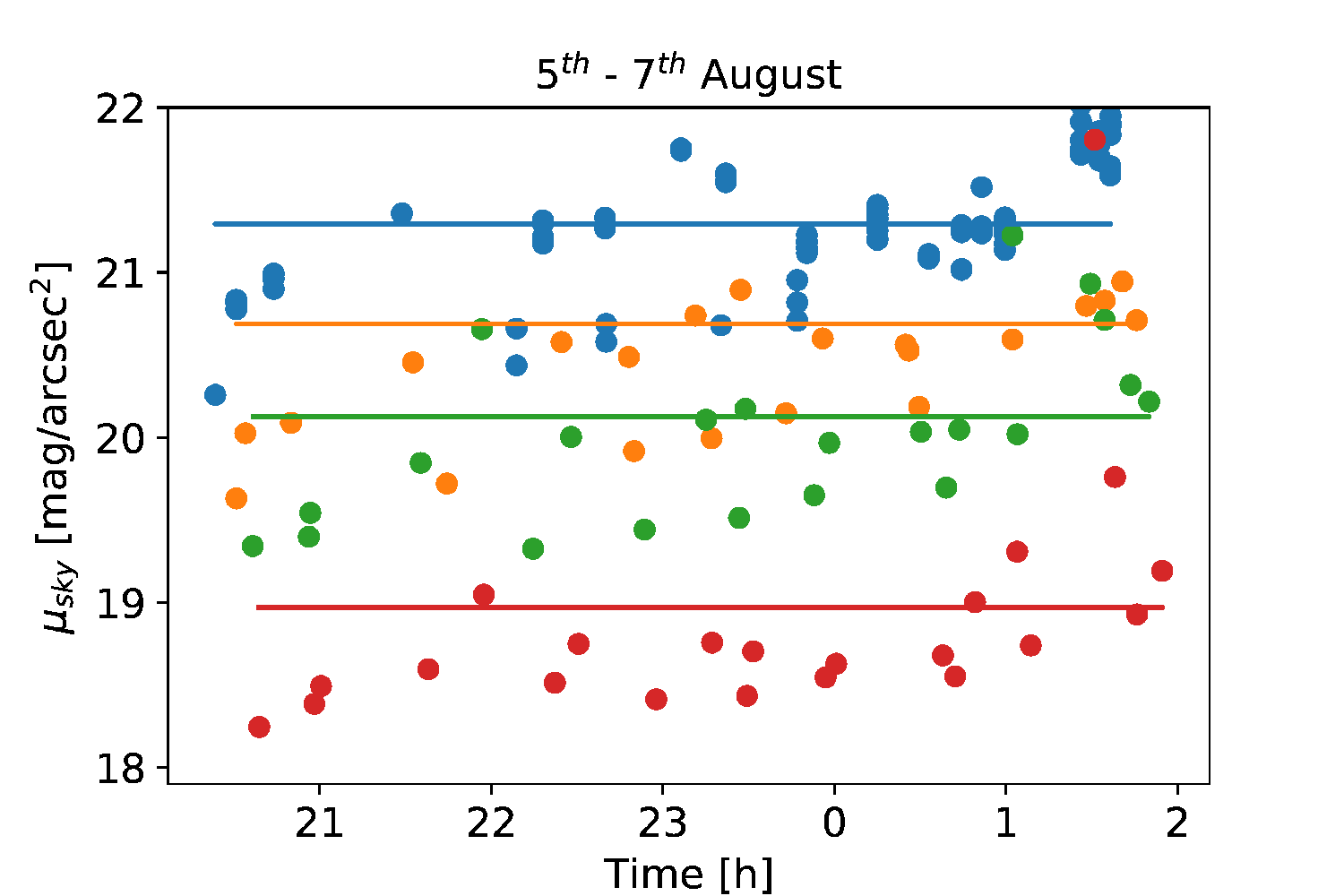}
\caption{Sky surface brightness in magnitudes per arcsec squared during the observing nights from 5\textsuperscript{th} to 7\textsuperscript{th} August, color coded according to the filter used: blue (B), orange (V), green (R) and  red (I). Horizontal lines -- colored according to the filter used -- mark the mean value of the sky brightness.}
\label{fig:sky}
\end{figure}

\begin{table}
\caption{Measured sky brightness in the magnitudes per arcsec squared in the $BVRI$ bands for three observing nights, from 5\textsuperscript{th} August to 7\textsuperscript{th} August.}
\vspace{.2cm}
\centerline{\begin{tabular}{|c|c|c|c|c|}
\hline
f& 5\textsuperscript{th} August & 6\textsuperscript{th} August & 7\textsuperscript{th} August \\ \hline \hline
$B$ & 21.29 $\pm$ 0.03 & 21.22 $\pm$ 0.08 & 19.85 $\pm$ 0.10 \\ \hline
$V$ & 20.86 $\pm$ 0.13 & 20.43 $\pm$ 0.15 & 19.39 $\pm$ 0.20 \\ \hline
$R$ & 20.17 $\pm$ 0.06 & 20.05 $\pm$ 0.10 & 19.18 $\pm$ 0.08 \\ \hline
$I$ & 18.77 $\pm$ 0.04 & 19.28 $\pm$ 0.21 & 18.42 $\pm$ 0.06 \\ \hline
\end{tabular}}
\label{tab:sky}
\end{table}

Measurements of the sky brightness are listed in the Table.~\ref{tab:sky} for each night. Photometric calibration was done using measurements from all three nights together (Fig.~\ref{fig:sky}), and in this case, when all the nights are taken together the measured sky brightness is: 20.86 $\pm$ 0.07, 20.32 $\pm$ 0.09, 19.86 $\pm$ 0.06, 18.81 $\pm$ 0.07 in the $BVRI$ bands, respectively (horizontal lines at Fig.~\ref{fig:sky}). Standard error of the mean sky brightness is similar to the variations in the calibrated magnitudes expressed as the standard deviation of the residuals between calibrated and standard magnitudes: 0.08, 0.06, 0.11 and 0.08 in the $BVRI$ bands (shaded area in the Fig.~\ref{fig:Bmulti} - \ref{fig:Imulti}).

The Moon was transiting from the New Moon phase (which was on the 1$^{st}$ of August, 2019) to the First Quarter, which was exactly on the 7$^{th}$ of August, 2019. Observing at this time  minimized the effect of the Moon light on the sky brightness. 

The time lapse videos from the All Sky camera installed at the ASV shows that the nights of the  5$^{th}$, 6$^{th}$ and 7$^{th}$  of August, 2019 were clear nights, without clouds\footnote{\url{https://www.youtube.com/watch?v=CrQTzXtBpcs}, \url{https://www.youtube.com/watch?v=R3aZPefB2MA} and \url{https://www.youtube.com/watch?v=BJUqi5-4XhY}} making them ideal for the photometric calibration observations.


\section{CONCLUSION}
\label{conclusion}
\indent

We have observed 31 field of standard stars in the $BVRI$ Johnson's photometric system, selected from the Landolt's catalog of standard stars with 60 cm Nedeljkovi\'c telescope equipped with FLI PL 230 CCD camera. Observations are conducted during three nights in August, 2019 at zenith distances $z<60^{\circ}$. We have done photometric calibration both with and without the color term in the $B$-, $V$-, $R$- and $I$-bands. Calibration coefficients measured by means of orthogonal distance regression are then used to transform instrumental to calibrated magnitudes. Simple extinction correction is sufficient in all the bands, except for the $B$-band, where additional, color correction need to be applied. Accuracy achieved inspecting magnitude errors that encompass both measurement and statistical errors added in the quadrature is between 2\% and 5\% for all the $BVRI$ bands.

\def\DJlatin{{\fontencoding{T1}\selectfont\char208}}
\acknowledgements{We acknowledge the financial support of the Ministry of Education, Science and Technological Development of the Republic of Serbia through the contract No.~451-03-9/2021-14/200002. We thank the Director of the Astronomical Observatory of Belgrade, Dr Gojko {\DJlatin}ura{\v{s}}evi{\'c}, for giving away his  observational time for this project, and the technical operators at the ASV, Miodrag Sekuli\'c and Petar Kosti\'c for their excellent work. This research made use of Photutils, an Astropy package for
detection and photometry of astronomical sources.}


\vskip2mm

\newcommand\eprint{in press }

\bibsep=0pt

\bibliographystyle{aa_url_saj}

{\small

\bibliography{calibration}
}

\clearpage

{\ }

\newpage

{\ }

\begin{strip}



\naslov{$BVRI$ FOTOMETRIJSKA KALIBRACIJA NEDELJKOVI{\CC} TELESKOPA}


\authors{A. Vudragovi{\' c}$^1$ and M. I. Jurkovic$^1$}

\vskip3mm


\address{$^1$Astronomical Observatory, Volgina 7, 11060 Belgrade, Serbia}


\Email{ana@aob.rs}

\vskip3mm


\centerline{{\rrm UDK} \udc}


\vskip1mm

\centerline{\rit Uredjivaqki prilog}

\vskip.7cm




\begin{multicols}{2}

{
\rrm

U ovom qlanku smo opisali fotometrijsku kalibraciju 60 $cm$  teleskopa Nede{\lj}kovi{\cc} sa ${\rm FLI\ PL\ 230\ CCD}$ kamerom, koji je postav{\lj}en na Astronomskoj stanici Vidojevica (Srbija), pomo{\cc}u Landoltovog kataloga standardnih zvezda. Napravili smo snimke 31.~polja standardnih zvezda koriste{\cc}i {\DZ}onsonove ${\rm BVRI}$ filtere, tokom tri no{\cc}i u avgustu 2019.~godine. Izraqunali smo korekciju na ekstinkciju i boju. Porede{\cc}i kalibrisane magnitude sa magnitudama standardnih zvezda iz Lanoltovog kataloga, postigli smo taqnost od 2\%-4\% za ${\rm BVRI}$ magnitude.
{\ }

}

\end{multicols}

\end{strip}


\end{document}